\documentclass[9pt,conference]{IEEEtran}
\IEEEoverridecommandlockouts

\usepackage{cite}
\usepackage{amsmath,amssymb,amsfonts}
\usepackage{algorithmic}
\usepackage{graphicx}
\usepackage{textcomp}
\usepackage{xcolor}

\usepackage{comment}
\usepackage{hyperref}

\let\forallsymb\forall

\input{mysymbol.sty}

\newcommand{\Tr}{\mathsf{T}}

\def\BibTeX{{\rm B\kern-.05em{\sc i\kern-.025em b}\kern-.08em
    T\kern-.1667em\lower.7ex\hbox{E}\kern-.125emX}}
\begin{document}

\title{Tracking Network Dynamics using Probabilistic State-Space Models
\thanks{Work partially supported by the EU H2020 Grant Tailor (No 952215, agreements 76 and 82), by the Spanish AEI (AEI/10.13039/501100011033), grants TED2021-130347B-I00, PID2023-149457OB-I00, PID2022-136887NB-I00 and FPU20/05554, the Community of Madrid via the Ellis Madrid Unit, the TU Delft AI Labs programme, the NWO OTP GraSPA proposal \#19497, and the NWO VENI proposal 222.032. Email contact author: antonio.garcia.marques@urjc.es.}
}

\author{Victor M. Tenorio$^{\dagger}$, Elvin Isufi$^{\star}$, Geert Leus$^{\star}$, Antonio G. Marques$^{\dagger}$ \\

$^{\dagger}$King Juan Carlos University, Madrid, Spain \{victor.tenorio,antonio.garcia.marques\}@urjc.es \\
$^{\star}$Delft University of Technology, Delft, The Netherlands \{e.isufi-1,g.j.t.leus\}@tudelft.nl
\vspace{-0.4cm}
}


\maketitle

\begin{abstract}
This paper introduces a probabilistic approach for tracking the dynamics of unweighted and directed graphs using state-space models (SSMs). Unlike conventional topology inference methods that assume static graphs and generate point-wise estimates, our method accounts for dynamic changes in the network structure over time. We model the network at each timestep as the state of the SSM, and use observations to update beliefs that quantify the probability of the network being in a particular state. Then, by considering the dynamics of transition and observation models through the update and prediction steps, respectively, the proposed method can incorporate the information of real-time graph signals into the beliefs. These beliefs provide a probability distribution of the network at each timestep, being able to provide both an estimate for the network and the uncertainty it entails.
Our approach is evaluated through experiments with synthetic and real-world networks. The results demonstrate that our method effectively estimates network states and accounts for the uncertainty in the data, outperforming traditional techniques such as recursive least squares.
\end{abstract}
\section{Introduction}
\label{sec:intro}

Nowadays, numerous algorithms and methods in data science leverage the underlying structure of the data by means of a graph~\cite{kolaczyk2009book,shuman2013emerging,djuric2018cooperative,ortega2018graph}.
However, most tools doing so assume that the graph is static and perfectly known, whereas in reality, that is rarely the case. Often, the graph must be inferred from the data~\cite{mateos2019connecting,das22connect}, exhibits perturbations~\cite{rey23rgfi,tenorio2023robust} or both the graph and the data defined over it vary with time~\cite{casteigts2012time,das24tensor}.
Therefore, the task of tracking network dynamics is crucial for the optimal performance of the downstream algorithms that leverage this network.
To provide a simple example, in air traffic, if we construct a graph where nodes represent airports and edges represent flights connecting them~\cite{yanagiya2024edge,tenorio21delay}, the number of flights between two airports fluctuates throughout the day. Thus, any architecture that assumes static edges between airports will likely create connections that do not always exist.

The importance of considering the time variation of the underlying topology is highlighted by the fact that there is an increasing body of literature dealing with dynamic graphs~\cite{ioannidis19blind,shafipour2019online}.
Overall, this requires assuming a particular model behavior between data and the evolving topology, such as Gaussian graphical models or structural equation models~\cite{natali22time}.
Another method that can be used in this setting, and one of the most widely used methods to deal with systems that vary over time in general, are state-space models (SSMs). SSMs represent the system with a state and track its evolution through time using observations, with one of the most prominent methods being the Kalman filter~\cite{kalman1960new}. SSMs, and specifically the Kalman filter, have been successful in a wide array of tasks, including sensor networks~\cite{shi90kalman}, navigation~\cite{grewal10aerospace} and many more.

The objective of this work is to use \emph{a probabilistic} SSM to track network dynamics, by considering the network as the state of the SSM and using graph (nodal) signals as observations.
Tracking dynamic graphs is a well-known problem within the GSP community, mainly solved by making assumptions about the observed graph signals, such as smoothness or stationarity~\cite{saboksayr21onlinesmooth,shafipour20onlinestat,cui24topinf,money23scalable}.
However, trying to solve this problem with SSMs is a hard problem that presents two main challenges: first, how to encode the network structure within the state, and second, how to incorporate observation data to update this state. 
Consequently, only a few studies have approached the problem by considering the network as the state of an SSM. Notable examples include~\cite{dabush24kalman}, which defines the state as the vector of edge weights and uses a sparsity-promoting algorithm based on the Kalman filter to track them, \cite{alippi2023graph,zambon2023graph} which consider the state and observations as attributed graphs and use a Kalman filter to track their evolution and \cite{shvydun2023system} which aims to find the temporal dynamics of the network by learning the system matrix, i.e., the matrix governing the SSM.
Instead of considering the network as the state of the SSM, most approaches encode the network in the matrices that govern the evolution of the SSM, such as~\cite{buchnik2023kalmannet,coutino2020state}.
In general, these methods offer several drawbacks compared to ours. The first one is that they provide a point estimate for the state at each timestep, whereas under a probabilistic setting, our method provides a probability distribution of the state at each timestep, which allows us to not only provide a point estimate of the state, but also the uncertainty underlying this estimate. The second one is that system identification is required to track the graph. This means that they require a large number of observations, and hence they are only able to track slowly varying graphs. Instead, by modeling the graph as the state of the SSM, our method is able to use filtering to track the dynamics of the network.

\noindent
\textbf{Contribution}: In this paper, we propose a novel method to track unweighted and directed network dynamics using a probabilistic SSM. We assume a probabilistic transition model and an observation model consisting of an exchange of signal information among the one-hop neighborhood of the graph, which allows us to model each row of the graph's adjacency matrix as a node-dependent state. Then, we track its evolution through time via the input-output graph signal pairs obtained from the observation model and related to the graph. The output of our model is the probability distribution of the state at each timestep, quantified by a belief vector. These beliefs allow us to quantify the uncertainty contained in the estimate of our method, being able to measure the confidence of the model on the provided estimation.



\section{Preliminaries}
\label{sec:prelim}

\noindent
\textbf{Graphs and graph signals}:
let $\ccalG = (\ccalV, \ccalE, \bbA)$ denote a graph, with $\ccalV$ representing the node set with cardinality $|\ccalV | = N$, $\ccalE \subset \ccalV \times \ccalV$ being its edge set and $\bbA$ being its adjacency matrix, with $[\bbA]_{nn'} \neq 0$ only if $(n',n) \in \ccalE$.
If the entries of $\bbA$ are binary, we are dealing with unweighted graphs.
Graphs are said to be undirected if $(n',n) \in \ccalE$ implies that $(n,n') \in \ccalE$ (equivalently, if $\bbA = \bbA^\Tr$) and are directed otherwise. Finally, we say that $\bbz\in\reals^N$ is a (nodal) graph signal if the $n$-th entry of $\bbz$ represents the value of a magnitude/feature associated with the $n$-th node of the graph. Graph signals can be analyzed and processed using graph-related operators such as the graph Fourier transform or graph filters~\cite{shuman2013emerging,isufi2024graph,tenorio24blind}.

\vspace{.05cm}
\noindent
\textbf{State-space models}: SSMs represent real-world systems by means of a state, defined by features collected in the vector $\bbx_t$ that evolves over time, and observations $\bby_t$ that are related to the state. The standard assumptions in SSMs are that: i) the dynamics are known and Markovian, so that $\bbx_t=g_t(\bbx_{t-1})$, with the transition function $g_t$ being possibly random; and ii) the observations are memoryless so that $\bby_t = f_t (\bbx_t)$, with $f_t$ denoting the (random) measurement function.
The goal of SSMs is to determine the evolution of the state over time, $\{\bbx_t\}_{t=1}^T$, given the sequence of observations $\{\bby_t\}_{t=1}^T$.

Probabilistic SSMs do not try to find an exact value for the state but rather a probability distribution or belief of the state at each timestep $t$. For example, when the transition and the observation functions are linear and the distributions involved are Gaussian, it is possible to track the mean vector and covariance matrix that describe the belief by means of a Kalman filter.
If the state space is discrete, i.e., there is a finite number of states, then it is possible to track the probability of being in each of the states at time $t$ through a belief vector $\bbb_t$, which collects the values of the probability mass function of the state. For the particular case of filtering, $\bbb_t = p(\bbx_t | \{\bby_\tau \}_{\tau = 1}^t )$. This last approach is the one adopted in this work.

\section{Problem Formulation}
\label{sec:prob}



This section presents the underlying problem we aim to solve, along with the assumptions considered for the models that govern the SSM. We first give a general description of the problem. Then we provide the considered transition and observation models. We finalize the section by providing the definition of the state and the formal problem definition.

This works deals with the problem of using a probabilistic SSM to track the evolution of a network represented by a directed and unweighted graph with $N$ nodes, whose topology at time $t$ is denoted by $\bbA_t$. As a result, in this paper the state $\bbx_t$ is represented by $\bbA_t$, which is an  $N\times N$ binary matrix. We further assume that: i) the graph is not directly observable; ii) at each time step $t$ we have access to signals defined on the nodes of the graph; and iii) the value of the observed signals depends on the topology encoded in $\bbA_t$. The goal is to use the existing observations up to time $t$ to estimate the probability mass function (belief) of $\bbA_t$.
To this end, we need a model for (i) the transition function that governs the evolution of the graph--transition model-- and (ii) the function that returns the observation by using the graph--observation model.

\vspace{.05cm}
\noindent
\textbf{Transition model.}
The transition model describes how the graph evolves over time. 
For simplicity, we consider a probabilistic entry-wise transition function, where the evolution of each edge is described by the probability mass function
\begin{equation}\label{E:transitionmodel_ij}
p([\bbA_t]_{nn'} | [\bbA_{t-1}]_{nn'}).    
\end{equation}
This transition model, described by a Markov process, assumes that the value of each edge depends only on its value in the previous timestep. This assumption of SSMs is prevalent in both the machine learning and signal processing communities~\cite{dabush24kalman,alippi2023graph,zambon2023graph,buchnik2023kalmannet}.

Since the graph is unweighted, both $[\bbA_{t-1}]_{ij}$ and $[\bbA_t]_{ij}$ are binary, so that four values, denoted as $P_{ij}^{00}$, $P_{ij}^{01}$, $P_{ij}^{10}$, and $P_{ij}^{11}$, suffice to describe the function in \eqref{E:transitionmodel_ij}. Using this notation, we have that
\begin{equation}\label{E:transitionmodel_ij_v2}
P_{ij}^{l l'}=p([\bbA_t]_{ij}=l| [\bbA_{t-1}]_{ij}=l').    
\end{equation}
For example, $P_{ij}^{01}$ represents the probability of link $(j,i)$ not being active in $\bbA_t$ given that it was present in $\bbA_{t-1}$.

\vspace{.05cm}
\noindent
\textbf{Observation model.}
The observation model considered is
\begin{equation}\label{eq:obs_model}
  \bby_t = \bbA_t \bbz_t + \bbw_t,
\end{equation}
where $\bbA_t\in\{0,1\}^{N\times N}$ is the GSO of the graph at timestep $t$, $\bbz_t\in\reals^N$ is a known input graph signal, $\bby_t\in\reals^N$ is an observed output graph signal, and $\bbw_t\in\reals^N$ is the observation noise. The key difference relative to most of the existing literature~\cite{sabbaqi2024InferringTV,zambon2023graph} is that, rather than assuming that we observe entries of the matrix $\bbA_t$, we assume that we only have access to signals defined on the graph. The model in \eqref{eq:obs_model} is relevant when there is a network-process diffusing information across the (one-hop neighborhood of the) graph, and can represent real-world interactions such as exchanges of information in social networks, one step of the propagation of epidemics, etc. \cite{segarra17system}. Two particular cases of \eqref{eq:obs_model} are i) autoregressive graph processes of order one~\cite{isufi2019forecasting}, where the input signal $\bbz_t$ corresponds to the previous observation, $\bbz_t = \bby_{t-1}$, and ii) sparse structural equation models~\cite{natali21onlineSEM}, where $\bbz_t = \bby_t$.


\vspace{.05cm}
\noindent
\textbf{State definition}. 
As stated, we aim to track the network dynamics of a directed and unweighted graph by modeling $\bbA_t$, the network at timestep $t$, as the state $\bbx_t$. Since the graph is unweighted and does not have self-loops, matrix $\bbA_t$ is binary and contains at most $N(N-1)$ non-zero entries. As a result, $\bbA_t$ can take at most $2^{N(N-1)}$ values and, hence, describing the belief $p(\bbA_t | \{\bby_\tau\}_{\tau=1}^t,\{\bbz_\tau\}_{\tau=1}^t)$ requires specifying the value of $2^{N(N-1)}$ probabilities, which is a large number even for small graphs.
In the next section, we will explain how to reduce the size of the state space to simplify the problem.

\vspace{.05cm}
\noindent
\textbf{Problem statement.} Given the models in \eqref{E:transitionmodel_ij}-\eqref{eq:obs_model}, the input signals $\{\bbz_\tau\}_{\tau=1}^t$ and the observations in $\{\bby_\tau\}_{\tau=1}^t$ estimate the probability mass function
\begin{equation}\label{eq:prob_mass_function_to_track_t}
p(\bbA_t | \{\bby_\tau\}_{\tau=1}^t,\{\bbz_\tau\}_{\tau=1}^t).
\end{equation}

After having introduced our problem, the next section designs and discusses the schemes to track the beliefs in \eqref{eq:prob_mass_function_to_track_t}.


\section{Online belief estimators} \label{sec:method}

We begin by describing how we represent the network at a given timestep as the state of the SSM, and then we detail the prediction and update steps used to update our beliefs.


\vspace{0.05cm}
\noindent
\textbf{Simplified state definition.}
As previously described, considering the whole matrix $\bbA_t$ as the state of our SSM yields a very large state space.
Interestingly, the model in \eqref{E:transitionmodel_ij} and \eqref{eq:obs_model} allows us to reduce the tracking complexity. Firstly, the dynamics in \eqref{E:transitionmodel_ij} are independent across links. Secondly, the observation model in \eqref{eq:obs_model} relates the $n$-th entry of $\bby_t$ with the $n$-th row of $\bbA_t$, meaning that \eqref{eq:obs_model} relates the observation at node $n$ with the links of the graph associated with the (incoming) neighbors of $n$. Building on these two properties, we can postulate a separate belief for each row of $\bbA_t$. Formally, let $[\bbA_t]_{n,:}$ be the the $n$-th row of $\bbA_t$. Since $[\bbA_t]_{nn}$ is zero (we assume there are no self-loops), the vector $[\bbA_t]_{n,:}$ can take at most $2^{N-1}$ different values and, as a result, the belief for $[\bbA_t]_{n,:}$ is described by $2^{N-1}$ probabilities. Since $\bbA_t$ has $N$ rows, the state of the full network is defined then by $N$ row beliefs, so that the number of parameters to track is $N2^{N-1}$, which is considerably smaller than $2^{N(N-1)}$.

Based on this preliminary discussion, we are ready to define the state of our probabilistic SSM.
To facilitate exposition, let us define $\ccalA^n \subset \{0, 1\}^{N}$ as the set of all possible binary vectors of length $N$ with a zero inserted in the $n$-th position. 
Furthermore, let us define $I=2^{N-1}$ and denote the $i$-th element of set $\ccalA^n$ as $\bba_i^n$, with $i=1,...,I$. According to our previous explanation, we consider $N$ different states $\{\bbx_t^n\}_{n=1}^N$, with $\bbx_t^n$ representing the values of the $n$-th row of $\bbA_t$. Since the SSM is probabilistic, we need to find the belief associated with each $\bbx_t^n$. And since $\bbx_t^n$ can take $I$ values, the belief is represented by a vector of dimension $I$. Mathematically, this entails considering two vectors $\bbb_{t|t-1}^n \in \reals^{I}$ and $\bbb_{t|t}^n \in \reals^{I}$ whose $i$-th entry is 
\begin{subequations} \label{eq:belief_def}
\begin{align}
    [\bbb_{t|t-1}^n]_i &= p(\bbx_t^n = \bba_i^n | \{\bby_\tau\}_{\tau=1}^{t-1},\{\bbz_\tau\}_{\tau=1}^{t-1}) \; \forallsymb \; t, n, \bba_i^n \in \ccalA^n,
    \label{eq:belief_def_apriori}\\
    [\bbb_{t|t}^n]_i &= p(\bbx_t^n = \bba_i^n | \{\bby_\tau\}_{\tau=1}^{t},\{\bbz_\tau\}_{\tau=1}^{t}) \; \forallsymb \; t, n, \bba_i^n \in \ccalA^n,  \label{eq:belief_def_aposteriori}
\end{align}    
\end{subequations}
where \eqref{eq:belief_def_apriori} represents the belief \textit{a priori} (i.e., right before observing $\bby_t$) and \eqref{eq:belief_def_aposteriori} represents the belief \textit{a posteriori}.

\vspace{0.05cm}
\noindent
\textit{Remark 1.}
Note that this row-dependent state definition is applicable to a transition model more general than the one presented in~\eqref{E:transitionmodel_ij}. Instead of considering that the value of each entry $[\bbA_t]_{nn'}$ depends only on its value in the previous timestep $[\bbA_{t-1}]_{nn'}$, we can consider dependencies on every other value of the same row $[\bbA_{t-1}]_{n,:}$. Thus, instead of being limited to the transition models whose probability mass function follows~\eqref{E:transitionmodel_ij}, our model can deal with transition functions described by $p([\bbA_t]_{n,:} | [\bbA_{t-1}]_{n,:})$ and handle a broader range of scenarios. 


\vspace{0.05cm}
\noindent
\textbf{Prediction step}.
The goal of the prediction step is to compute the \textit{a priori} beliefs $\bbb_{t|t-1}^n$ from the observations up until time $t-1$ and the prior information about the transition model. Thanks to Markovianity, this can be carried out by the transition dynamics in \eqref{E:transitionmodel_ij} plus $\bbb_{t-1|t-1}^n$, the \textit{a posteriori} beliefs computed at time $t-1$. More precisely, we compute the \textit{a priori} belief for state $i$ and node $n$ using the law of total probability as indicated in equation \eqref{eq:prediction_belief_step_full}. 
%
%
In \eqref{eq:prediction_belief_step_full}, the \textit{a posteriori} belief in the previous timestep $\bbb_{t-1|t-1}^n$ represents the probability of being in state $\bba_j^n \in \ccalA^n$ at timestep $t-1$, and $p( \bbx_t^n = \bba_i^n | \bbx_{t-1}^n = \bba_j^n, \{\bby_\tau\}_{\tau=1}^{t-1},\{\bbz_\tau\}_{\tau=1}^{t-1})$ incorporates the available information of the state transition model. This expression can be compactly written as a matrix-vector multiplication given by
\begin{equation}
    \bbb_{t|t-1}^n = \bbF_t^n \bbb_{t-1|t-1}^n,
\end{equation}
where the matrix $\bbF_t^n \in [0,1]^{I \times I}$ is the transition matrix for node $n$ with entries $[\bbF_t^n]_{ij} = p( \bbx_t^n = \bba_i^n | \bbx_{t-1}^n = \bba_j^n, \{\bby_\tau\}_{\tau=1}^{t-1},\{\bbz_\tau\}_{\tau=1}^{t-1})$.

\vspace{0.05cm}
\noindent
\textbf{Update step}.
After computing the \textit{a priori} beliefs, the update step integrates the information from the observation $\bby_t$ and the observation model using Bayes' rule as indicated in equation \eqref{eq:update_belief_step_full}.
\begin{table*}
\begin{center}
\begin{align}
    [\bbb_{t|t-1}^n]_i &= \sum_{j=1}^{I} p( \bbx_t^n = \bba_i^n | \bbx_{t-1}^n = \bba_j^n, \{\bby_\tau\}_{\tau=1}^{t-1},\{\bbz_\tau\}_{\tau=1}^{t-1}) p(\bbx_{t-1}^n = \bba_j^n|\{\bby_\tau\}_{\tau=1}^{t-1},\{\bbz_\tau\}_{\tau=1}^{t-1}) \nonumber \\
    &= \sum_{j=1}^{I} p( \bbx_t^n = \bba_i^n, \{\bby_\tau\}_{\tau=1}^{t-1},\{\bbz_\tau\}_{\tau=1}^{t-1} | \bbx_{t-1}^n = \bba_j^n) [\bbb_{t-1|t-1}^n]_j,\label{eq:prediction_belief_step_full} \\
\label{eq:update_belief_step_full}
    [\bbb_{t|t}^n]_i &= \frac{p([\bby_t]_n | \bbx_t^n = \bba_i^n, \{\bby_\tau\}_{\tau=1}^{t-1},\{\bbz_\tau\}_{\tau=1}^{t}) p(\bbx_t^n = \bba_i^n|\{\bby_\tau\}_{\tau=1}^{t-1},\{\bbz_\tau\}_{\tau=1}^{t-1})}{p([\bby_t]_n|\{\bby_\tau\}_{\tau=1}^{t-1},\{\bbz_\tau\}_{\tau=1}^{t})} = \frac{p([\bby_t]_n | \bbx_t^n = \bba_i^n, \{\bby_\tau\}_{\tau=1}^{t-1},\{\bbz_\tau\}_{\tau=1}^{t}) [\bbb_{t|t-1}^n]_i}{p([\bby_t]_n|\{\bby_\tau\}_{\tau=1}^{t-1},\{\bbz_\tau\}_{\tau=1}^{t})}.
\end{align}
\noindent\rule{\linewidth}{0.4pt}
\vspace{-.1cm}
\end{center}
\end{table*}
Note that the right-most expression in \eqref{eq:update_belief_step_full} considers  three components: i) the likelihood of the observation given the state and the known signal input (first term in the numerator), ii) the \textit{a priori} belief on the state (second term in the numerator), and iii) the likelihood of the observation (denominator). The likelihood $p([\bby_t]_n | \bba_i^n, \{\bby_\tau\}_{\tau=1}^{t-1},\{\bbz_\tau\}_{\tau=1}^{t})$ is derived from the observation model.
For example, with Gaussian observation noise $\bbw_t \sim \ccalN(\bbzero, \sigma_{obs}^2 \bbI)$, each entry of the output is distributed as
\begin{equation}\label{eq:y_dist_gauss}
 p([\bby_t]_n | \bbx_t^n =\bba_i^n, \{\bby_\tau\}_{\tau=1}^{t-1},\{\bbz_\tau\}_{\tau=1}^{t-1})= \ccalN (\bba_i^{n\;\Tr} \bbz_t, \sigma_{obs}^2).  
\end{equation}
%
Using the likelihood given the state, the \textit{a priori} beliefs and the law of total probability, the likelihood in the denominator $p([\bby_t]_n|\{\bby_\tau\}_{\tau=1}^{t-1},\{\bbz_\tau\}_{\tau=1}^{t})$ is computed as
\begin{equation}
     \sum_{j=1}^{I} p([\bby_t]_n | \bbx_t^n = \bba_i^n, \{\bby_\tau\}_{\tau=1}^{t-1},\{\bbz_\tau\}_{\tau=1}^{t}) [\bbb_{t|t-1}^n]_i.
\end{equation}
In words, the update step in \eqref{eq:update_belief_step_full} re-scales the values in the \emph{a priori} belief with weights given by the likelihood obtained from the observation model and then re-normalizes the resulting vector, so that the aggregated probability is one. 

In summary, our model comprises 2 steps: first, the prediction step, which updates the \textit{a priori} beliefs through a matrix-vector multiplication based on the transition model. Second, the update step incorporates the information about the observations to compute the \textit{a posteriori} beliefs. This approach provides probabilistic estimates of the network at each timestep, offering not only the network estimate but also the associated uncertainty. 

\vspace{0.05cm}
\noindent
\textit{Remark 2.}
The complexity of the prediction step of our algorithm is $\ccalO (I^2 N)$, as it implies $N$ matrix-vector multiplications of size $I$. The update step carries a complexity of $\ccalO (3 I N)$ as it implies computing the conditional likelihood $I$ times, $I$ multiplications and $I$ additions, all of this repeated $N$ times. Therefore, if we run the scheme for $T$ iterations the overall computational complexity of our algorithm is $\ccalO (I^2 NT)$. The journal version of this paper will further discuss (approximate) methods to reduce this complexity.


\section{Numerical Experiments}
\label{sec:exps}

This section presents numerical experiments to demonstrate the benefits of our probabilistic algorithm with both synthetic and real-world graphs. The code used to run the experiments can be found in \url{https://github.com/vmtenorio/prob-ssm}


\begin{figure*}
  \begin{minipage}[b]{0.48\textwidth}
    \centering
    \centerline{\includegraphics[width=8cm]{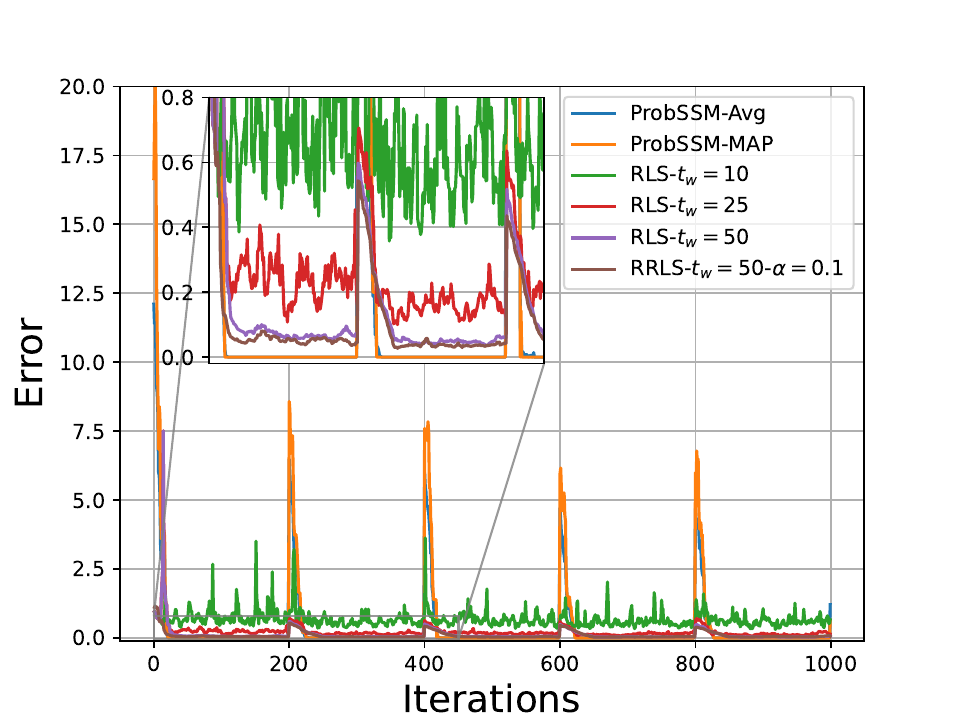}}
    \vspace{-6cm}
    \centerline{\large{(a) Synthetic data}}
    \vspace{5.5cm}
  \end{minipage}
  \begin{minipage}[b]{0.48\textwidth}
    \centering
    \centerline{\includegraphics[width=8cm]{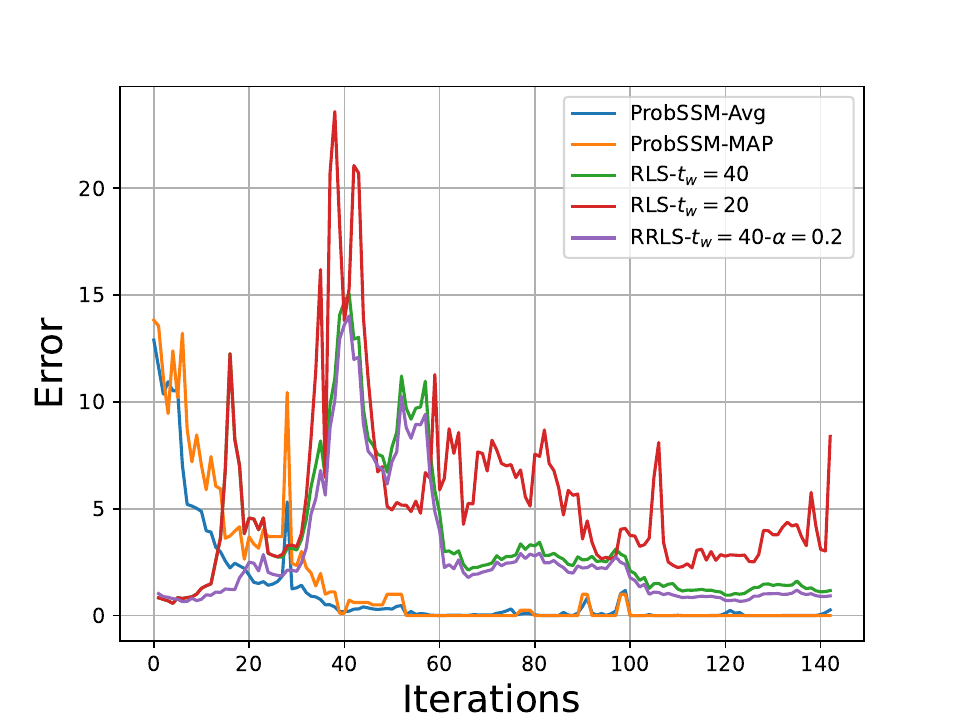}}
    \vspace{-6cm}
    \centerline{\large{(b) Airports data}}
    \vspace{5.5cm}
  \end{minipage}
  \caption{Performance of the proposed method in the two different experiments proposed: (a) contains the plot of the state error $\ccalL_t$ vs time $t$ for the synthetic data experiment presented in Section~\ref{sub:synth} and (b) contains the state error vs time for the airport graph of the experiment in Section~\ref{sub:airport}.}
  \vspace{-0.01cm}
  \label{fig:results}
\end{figure*}

\subsection{Synthetic Data} \label{sub:synth}

In the first experiment, we work with synthetically generated graphs and data samples. The graphs are generated using the Erd\H{o}s-Rényi random graph model with $N = 14$ nodes and a connection probability of $p = 0.25$~\cite{kolaczyk2009book}. To gain stronger insights, we consider a variation of the dynamics in~\eqref{E:transitionmodel_ij} where every $T_c = 200$ timesteps, each entry of the binary adjacency matrix has a probability $p_c = 0.2$ to flip its value from 0 to 1 or vice versa, reflecting the creation and destruction of edges in the unweighted and directed graph. Given this transition model, the $(i,j)$-th entry of the transition matrix $\bbF_t^n$ is 
\begin{equation}
[\bbF_t^n]_{ij} = 
\begin{cases}
[\bbI]_{ij}, & \text{if } t \; \text{mod} \; T_c \neq 0, \\
p_c^{d_{ij}^n} \cdot (1-p_c)^{N-1-d_{ij}^n}, & \text{if } t \; \text{mod} \; T_c = 0,
\end{cases}
\end{equation}
where $d_{ij}^n$ is the Hamming distance between states $\bba_i^n$ and $\bba_j^n$.
The observation model follows~\eqref{eq:obs_model}, with noise $\bbw_t \sim \ccalN(\bbzero, \sigma_{obs}^2 \bbI)$, such that the likelihood of each entry of the output $[\bby_t]_n$ is given by~\eqref{eq:y_dist_gauss}. For simplicity, the input vectors $\bbz_t$ are generated as independent realizations of $\ccalN (\bbzero, \bbI)$.


The performance of our method is measured using the normalized error of the predicted state as 
\begin{equation}
  \ccalL_t = \sum_{n=1}^N \|\hba_t^n - [\bbA_t]_{n,:}\|_2^2 / \|[\bbA_t]_{n,:}\|_2^2,
\end{equation}
where $\hba_t^n$ represents the estimate of the state of node $n$ at timestep $t$. This estimate can be calculated either as the expected state $\hba_t^n = \sum_{i=1}^{I} [\bbb_{t|t}^n]_i \bba_i^n$ (``Avg'' in the figure legends) or using the MAP estimate $i_{max} = \text{arg} \max_i [\bbb_{t|t}^n]_i$ and $\hba_t^n = \bba_{i_{max}}^n$ (``MAP'' in the figure legends).

We compare our algorithm against recursive least squares (RLS), whose output is computed as the least squares estimate of $\bbA_t$ using a rolling window of size $t_w$, by solving 
\begin{equation}
  \hbA_t=\arg\min_{\bbA} \| \bbY_t - \bbA \bbZ_t \|_F^2 + \alpha h(\bbA)
  \label{eq:rls}
\end{equation}
where $\bbY_t$ and $\bbZ_t$ $\in \reals^{N \times t_w}$ contain the output and input observations from timestep $t-t_w$ to $t$, respectively, and $h(\bbA)$ represents a regularization term that in the experiments is set to the sparsity-promoting $\ell_1$ norm yielding the Lasso estimate of $\bbA_t$. The estimate in~\eqref{eq:rls} setting $\alpha=0$ is labeled ``RLS'' in the figures, with ``RRLS'' representing the regularized version using the $\ell_1$ norm.


Figure~\ref{fig:results}-(a) illustrates the normalized error of our method compared to RLS over $T=1000$ timesteps, shown in the $x$-axis.
Our method can reach a negligible level of error in estimating the state, whereas RLS consistently has a higher error regardless of the window size.
Moreover, our method recovers more quickly from network changes, promptly returning to an accurate state estimate, while RLS requires $t_w$ samples to return to its lowest error value after a network change (since every sample in $\bbY_t$ must be computed using the updated $\bbA_t$).
This is caused by the fact that, as we use the prior information of the transition model, and then incorporate the observation information into the belief, we are able to quickly transition to the new state of the network.
The performance of RLS worsens when $t_w < N$ (e.g., green line with $t_w = 10$), and although increasing the window size reduces the error, it also necessitates more samples to recover from a network change.
Finally, the performance of both the expected and MAP versions of our algorithm is similar, with almost overlapping error curves in the figure.

\subsection{Airports Data} \label{sub:airport}

The final experiment utilizes a graph whose nodes are the $N=16$ biggest airports in Europe based on the number of daily flights, with edges existing between two airports if there are more than 4 daily flights between them.
The network dynamics are modeled such that, at each timestep $t$, there is a probability $p_e = 0.1$ that an airport closes, which would remove all its incoming edges (set them to 0).
The signal $\bbz_t$ is the number of flights within a window of 2 hours centered at time $t$ as collected from~\cite{oagData}.
The observation model in~\eqref{eq:obs_model} can be interpreted as air traffic flowing through the network, after the movements due to the flights happenning in the interval between $t-1$ and $t$. The synthetically generated noise term $\bbw_t \sim \ccalN (\bbzero, \sigma_{obs}^2 \bbI)$ aims to represent the effect of flight delays (negative noise for delayed flights that should have arrived at time $t$ but they arrive late, positive noise for delayed flights arriving later than planned).

Figure~\ref{fig:results}-(c) reports the results for this experiment, specifically showing the evolution of the normalized state error $\ccalL_t$ over time. When an airport closes, an entire row of $\bbA_t$ becomes zero, causing instability in the RLS method until the corresponding sample is out of the window and making the algorithm struggle to return to an error of 0. In contrast, our algorithm quickly and consistently recovers from such disruptions, effectively estimating the network within just a few iterations after the event.


\section{Conclusions}
\label{sec:conclusions}

This paper introduces a novel probabilistic algorithm for tracking the dynamics of directed, unweighted networks, leveraging nodal signal observations to estimate the network's state while quantifying the uncertainty of these estimates. Our approach is grounded in a SSM whose state is the network, and where we combine prior information of the transition model with observations in the form of nodal signals, offering a more comprehensive framework for network tracking. This integration represents a significant advancement over existing methods, which often overlook this dual layer of information. Through a series of experiments, both with synthetic and real-world data, we demonstrate that our method outperforms traditional approaches like RLS. Our algorithm's ability to model uncertainty and provide robust estimates in the face of incomplete or noisy data underscores its potential for applications in real-world network analysis and monitoring tasks.

\bibliographystyle{IEEEbib}
\bibliography{biblio}

\end{document}